\documentclass[aip,reprint]{revtex4-1}
\usepackage{epsfig,dsfont,amssymb,amsmath,amsthm,amsfonts,amsbsy,mathrsfs}

\usepackage{amsthm}
\usepackage{amssymb}
\draft 
\usepackage{color}
\usepackage[colorlinks, citecolor=blue]{hyperref}
\usepackage[figuresright]{rotating}
\usepackage{graphicx}

\begin{document}


\title{Ferroelectric field effect of the bulk heterojunction in polymer solar cells}



\author{Meng Li}
\author{Heng Ma}\altaffiliation{Electronic mail:hengma@henannu.edu.cn.}
\author{Hairui Liu}
\author{Yurong Jiang}


\affiliation{College of Physics \& Electronic Engineering, Henan Normal University, Xinxiang, 453007, PR China}
\affiliation{Henan Key Laboratory of Photovoltaic Materials, Xinxiang, 453007, PR China}
\author{Heying Niu}
\author{Adil Amat}

\affiliation{College of Physics \& Electronic Engineering, Henan Normal University, Xinxiang, 453007, PR China}


\date{\today}

\begin{abstract}
A ferroelectric field effect in the bulk heterojunction was found when an external electric field (EEF) was applied on the active layer of polymer solar cells (PSCs) during the annealing process of the active layer spin-coated with poly (3-hexylthiophene):[6,6]-phenyl-C$_{61}$ butyric acid methyl ester (P3HT:PCBM). For one direction field, the short circuit current density of PSCs was improved from 7.2 to 8.0 mA/cm$^{2}$, the power conversion efficiency increased from 2.4 to 2.8\textsf{\%}, and the incident photon-to-current conversion efficiency increased from 42 to 49\textsf{\%} corresponding to the different EEF magnitude. For an opposite direction field, the applied EEF brought a minus effect on the performance mentioned above. EEF treatment can orientate molecular ordering of the polymer, and change the morphology of the active layer. The authors suggest a explanation that the ferroelectric field has been built in the active layer, and therefore it plays a key role in PSCs system. A needle-like surface morphology of the active film was also discussed.
\end{abstract}


\maketitle


Polymer solar cells (PSCs) have attracted considerable attention due to their light weight, flexibility, low cost, and so on \cite{krebs2004production,krebs2009complete,sondergaard2012roll,krebs2009roll}. Recently, PSCs with power conversion efficiency (PCE) of 6-8\textsf{\%} have been achieved, based on conventional device architecture, using advanced conjugated polymers as donors and fullerene derivatives as acceptors \cite{li2012polymer,liang2010bright,peters2012mechanism,chen2009recent}. However, compared with Si-based solar cells their PCE is quite low, and the stability and lifetime quite poor, which has limited the industrialization of PSCs. Consequently, further improvement in PSC device performance is necessary.
The most commonly investigated PSCs produced from solution are comprised of the conjugated polymers poly-3(hexyl-thiophene) (P3HT) serving as a donor, and [6, 6]-phenyl-C$_{61}$ butyric acid methyl ester (PCBM) as an acceptor \cite{dennler2009polymer,li2005high,dyakonov2004mechanisms}. Many approaches have been used to improve the performance of the P3HT: PCBM system, such as thermal annealing, solvent annealing and the addition of additives \cite{li2005investigation,li2007solvent,lee2008processing,kim2013analysis}. These processes mainly focus on further increasing the molecular crystallinity, improving the orderliness of the molecular arrangement, and optimizing the morphology of the active layer \cite{liu2012highly,pearson2012rationalizing,kastner2012morphology}. Ultimately these approaches are capable of enhancing the carrier mobility and improving charge transport; however, they cannot prevent the active layer materials, especially PCBM molecules, from cluster aggregating which leads to a non-uniformity and a disorder of bulk heterojunction, thereby influencing the carrier mobility and charge transport \cite{tang2010precise}. This is the key factor in the subsequent poor stability and lower PCE of PSCs.

In order to optimize the morphology of the active layer, and reduce PCBM aggregation, we introduces a method to manipulate the surface and the internal of the organic molecular by applying external electric field (EEF) treatment during annealing of the active layer. Because applying electric field treatment during thermal annealing can achieve more optimal phase separation of the active layer, while maintaining mobility of the internal molecules, application of an EEF during this stage may effectively improve uniformity and order in the formation of a bulk heterojunction. In this paper, different EEFs were applied during annealing of the active layer; the influence of this treatment on PSCs performance and the physics of this phenomenon being subsequently investigated based on the polarity of organic polymer materials and morphology of the active layer.

A pre-patterned indium tin oxide (ITO) glass with a sheet resistance of average 15 $\Omega/\Box$ and transmission 85\textsf{\%} was purchased from Kaivo. Regioregular active materials P3HT and PCBM were purchased from Luminescence Technology Corp. and Solenne BV. As an aqueous dispersion, Poly (3, 4-ethy-lenedioxythiophene):poly (4-styrenesulfonate) (PEDOT: PSS) (AI 4083) was purchased from Bayer AG. All the chemicals were used as received without further purification.

The structure of the device, Glass / ITO / PEDOT:PSS (40nm) / P3HT: PCBM (150nm) / LiF (0.8nm) / Al (100nm), is shown in Fig. 1(a). The molecular structure attached its direction of dipole moment (4.18D) of PCBM, which is optimized with a semi empirical program is sketched in Fig. 1(b). In the annealing process of active layer, a vertical EEF was exerted on the blended layer of P3HT:PCBM using another ITO substrate cover. As the annealing temperature decreases, the applied EEF was kept constant till the temperature drop down to room temperature. The diagram of the device exerted the electric field is shown in Fig. 1(a). The treatments were divided into two conditions according to the direction of EEF: up $(+)$ and down $(-)$ electric field. A polarized PCBM figure within the active layer induced by EEF was highlighted in the right side.

All of the production processes adopted in this work are listed as follow.
PEDOT: PSS spin coating (thickness: 40 nm): 2500 rpm 15 s / 3000 rpm 45 s / 140$^{\circ}\mathrm{C}$ annealed 30 min (in air).
Preparation of active layer solution: P3HT 10 mg, PCBM 9 mg, 1, 2-dichlorobenzene 1 mL are mixed and stirred at 50$^{\circ}\mathrm{C}$ for 24 h.
Spin coating of active layer (thickness 150 nm): 800 rpm 1 min / 110$^{\circ}\mathrm{C}$ annealed 10 min (in air).
Al electrode thickness: 100 nm.
Solar simulator: 1000 W Xenon lamp simulator source AM 1.5 G, 100 mW/cm$^{2}$ (Abet Technologies Cop.).
Current density voltage (J-V) curves measurement: four wires method, Keithley 2400 source meter.
Atomic force microscopy (AFM) measurement: tapping mode (Veeco NanoScope 3D).
Incident photon-to-current conversion efficiency (IPCE) measurement: 1000 W halogen lamp, grating monochromator (Acton Spectra Pro 2300i).

To clearly depict the measured J-V characteristics of the devices, Fig. 2(a) shows a portion of J-V data with and without EEF. The key cell parameters including short-circuit current density (J$_{sc}$), Open circuit voltage (V$_{oc}$), Fill Factor (FF), PCE, series resistor (R$_{s}$) and shunt resistor (R$_{sh}$) are listed in Table 1. For a device with both up and down EEF treatments, the obvious difference in the J$_{sc}$ data is interesting. Compared without EEF treatment, EEFs with an up direction resulted in a marked increase of J$_{sc}$, and furthermore, J$_{sc}$ increased successively as the magnitude of the up EEF increased. Conversely, the down EEF brought about a slight decrease in J$_{sc}$ compared with no EEF treatment. From Table 1, the difference in J$_{sc}$ between the largest up EEF (+2.1$\times$10$^{6}$ V/m), and the largest down EEF, reaches 1.5 mA/cm$^{2}$. For V$_{oc}$ the differences are insignificant with variation in the EEF, regardless of the up or down direction. It is obvious that PCEs show a strong correlation with the magnitude and direction of EEF. For up EEF, PCEs increase with increasing EEF intensity almost linearly until EEF being 1.7$\times$10$^{6}$ V/m. From 2.1$\times$10$^{6}$ V/m, an approaching value of the breakdown field strength, the increase slop becomes gentile, indicating an ultimate value. On the contrary, PCEs decreases significantly with increasing down EEF in a nonlinear correlation till EEF being -2.1$\times$10$^{6}$ V/m. Varying from 2.4\textsf{\%} (no EEF) to 2.8\textsf{\%} (+2.1$\times$10$^{6}$ V/m), a maximum increase in PCE of 16\textsf{\%} was obtained with the up EEF. For down EEF, PCEs reduce from 2.4\textsf{\%} to 2.1\textsf{\%} (-2.1$\times$10$^{6}$ V/m) with and without EEF.

To identify the mechanism for increasing the photocurrent in PSCs, the IPCE was also measured \cite{pearson2012rationalizing,bai2013polymer} and shown in Fig. 2(b). IPCE depends on not only the light absorption, but also the charge collection which relates the exciton diffusion, separation and carrier transport. The applied EEFs did not appear to change the absorption spectra domain compared to the standard PSC, but can increase or reduce IPCE. For the up EEFs, the maximum value of IPCE is increased to 49\textsf{\%} which expresses a notable improvement over the maximum IPCE of 42\textsf{\%} compared to the standard cell. For the down EEF, the maximum IPCE was reduced to 36\textsf{\%}. It indicates that the different direction of EEF treatment can strongly influence the light absorption and the exciton diffusion, separation and carrier transport.

It is worth noting that the active materials, P3HT and PCBM, are both polar molecules. Using a semi-empirical method and Gaussian 09 software \cite{chiechi2013modern}, we calculated and obtained the dipole moments of P3HT and PCBM as 0.148 D and 4.18 D, respectively. The results show that the moment direction of PCBM molecule is parallel to its side chain as shown in Fig. 1(b). With the applied EEF, the strong polar molecule of PCBM is almost certainly to exhibit a strong polarization, and result in a more uniform ordering heterojunction within the active layer.

Because the applied EEF was kept constant till the temperature drop down to room temperature in the annealing process, the strong polar molecules of PCBM are ordered by EEF at annealing temperature. The temperature lower, the layer crystal formed is. Therefore, an inside field named ferroelectric polar field (FPF)\cite{Ren2014modern} induced by EEF within the active layer is built. As shown in Fig. 1(a), for the down EEFs, the positive polarization charge layer stacks toward to PEDOT:PSS layer. The inside FPFs of the active layer make the holes which are generated from the separate excitons deviate from the hole-collecting layer PEDOT:PSS and ITO. Therefore, the collecting efficiency of the positive carriers will be reduced naturally. On the contrary, the up EEF form a down FPF towards to the hole-collecting layer, this effect is beneficial for holes to drift towards to the hole-collecting layer PEDOT:PSS. As a conclusion, the up EEF treatment on the active layer can effectively enhance carriers transport; however, the down EEF treatment is just the opposite.

In order to confirm the polarization effect on the photocurrent, The stability of the unwrapped cells in the air with different electric field treatment was measured. The result of PCEs as a function of cell life times is shown in Fig. 2(c). In Fig. 2(c), a normalized process is performed on PCE values, where the largest PCE of the cell treated by the largest EEF (-2.1$\times$10$^{6}$ V/m) is set as 1, and the others are divided by the largest PCE. One can find that, all the values of PCEs are decay, and reduced with the testing times (20min, 2h, 6h, 12h, 24h). However, the reducing tendency is dependent with EEF treatment, i.e. the reducing gradient is approximate equivalent. They remain a strong relatively stable with the EEF treatment. This illustrates that the effect is not due to the space charge accumulation on the electrode/film interface. Therefore, the equal gradient decay of the cells indicates that molecular polarization of the active layer caused by EEF treatments results in the difference of PCEs.

The morphology of the thin film surface is another key factor for bulk heterojunction in PSCs. Fig. 3 shows AFM images of the surface morphology of the film produced from P3HT:PCBM under different EEF treatments. The films change from a smooth and bland morphology to rough and tussocky one with increasing strength of EEF. This indicates that the applied electric field induces polar molecules, in particular PCBM, to form a spicular surface during the annealing process. The measured surface roughness (root-mean-square, RMS) for films with and without electric fields (up direction) is as follows.

0.76 nm (0 V/m), 1.42 nm (+1.7$\times$10$^{6}$ V/m), 1.62 nm (+2.1$\times$10$^{6}$ V/m), 1.10 nm (-1.7$\times$10$^{6}$ V/m), 1.32 nm (-2.1$\times$10$^{6}$ V/m).

Generally, an increasing of RMS of the active layer can increase contact areas of electrodes and the active layer. This can lead to a lower series resistor, and subsequently a larger short-circuit current density J$_{sc}$ in PSCs. The experiments show that, no matter for up and down EEF treatment, the RMS increases with the strength of the electric field. However, the devices with down EEF treatment has not improved the current density compared with no EEF treatment, furthermore with the up EEF treatment. This suggests that there is another factor influencing the PCE of the cell. This result show from a different angle that the polarization effect does exist and play a key role in improving the performance of the devices.

The ferroelectric field effect in the bulk heterojunction was firstly proposed in processing the polymer solar cells made of P3HT:PCBM. According to the different direction of inside ferroelectric field induced by EEF, the influence on carrier transport is radically distinct. The application of down EEF on devices was found to result in a more orderly arrangement of P3HT and PCBM molecules in the active layer of PSCs. Meanwhile, AFM analysis revealed that EEF can cause the film surface morphology to become more needle-like. This morphology provides a larger contact area between the active layer and an Al electrode, thereby improving the carrier mobility. The experimental results show that treatment with up electric fields can increase the J$_{sc}$ and PCE of P3HT:PCBM polymer solar cells; however, it has minimal effect on V$_{oc}$. It is expected that this work will be of benefit to optimizing the performance of PSCs not only in laboratory research, but also in commercial applications.

This work is supported by the National Natural Science Foundation of China under grant No. 11074066.

%

\newpage

\begin{figure*}[htbp]
\centering
\includegraphics[width=0.8\textwidth]{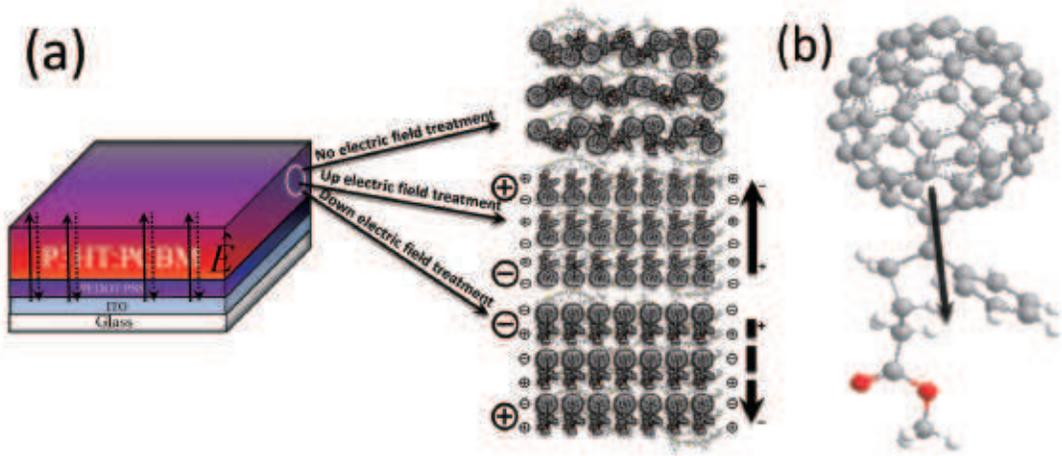}

\caption{(a) Schematic structure of the direction of EEF, which the solid arrow is defined as up (+), and the dash arrow as down (-).The polarization schematic of PCBM molecules under an EEF. (b) Structure of PCBM with  the direction of the dipole moment.}
\label{F1}       
\end{figure*}

\begin{figure*}[htbp]
\includegraphics[width=0.5\textwidth]{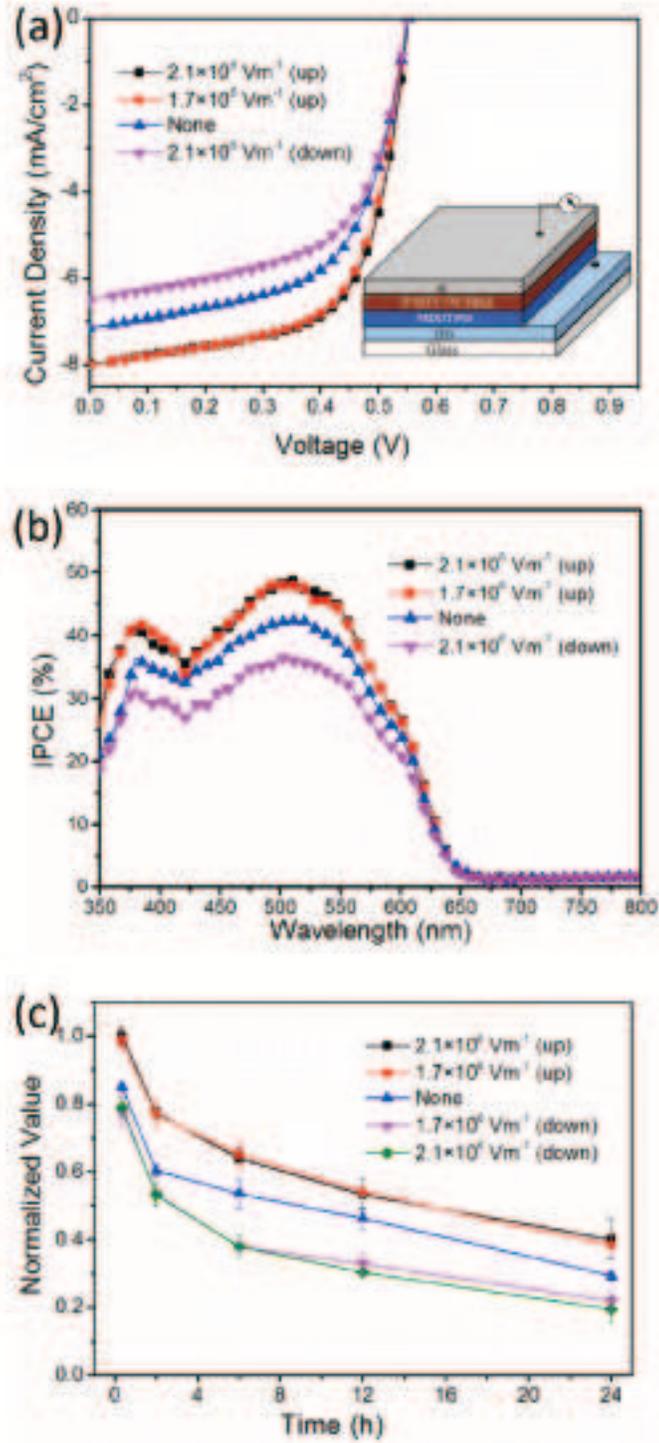}

\caption{(a) Schematic structure of PSCs and J-V characteristics with and without EEF treatment (+2.1$\times$10$^{6}$ V/m, +1.7$\times$10$^{6}$ V/m, -2.1$\times$10$^{6}$ V/m), (b) IPCE of PSCs as a function of wavelength under different EEF treatment, (c) The stabilities of Normalized PCE with different EEFs treatment (+2.1$\times$10$^{6}$ V/m, +1.7$\times$10$^{6}$ V/m, -1.7$\times$10$^{6}$ V/m, -2.1$\times$10$^{6}$ V/m).}
\label{F2}       
\end{figure*}

\begin{table*}[htbp]
\caption{The performance of P3HT: PCBM PSCs under different EEF treatment.}
\label{tab:1}       
\centering
\begin{tabular}{ccccccc}
\noalign{\smallskip}\hline
\hline\noalign{\smallskip}
\textrm{External electric field (Vm$^{-1}$)}
& $\emph{\textrm{J}}_{\textrm{sc}}(\textrm{mA}/\textrm{cm}^{2})$
&$\emph{\textrm{V}}_{\textrm{oc}}(\textrm{\textrm{V}})$
& \textrm{FF} & $\textrm{PCE}(\textsf{\%})$
& $\emph{\textrm{R}}_{\textrm{s}}(\textrm{$\Omega$}\textrm{cm}^{2})$
& $\emph{\textrm{R}}_{\textrm{sh}}(\textrm{$\Omega$}\textrm{cm}^{2})$\\

\noalign{\smallskip}\hline\noalign{\smallskip}
2.1$\times$10$^{6}$ V/m (up)  &7.99&0.55&0.64&2.80&11&504\\
1.7$\times$10$^{6}$ V/m (up)  &8.00&0.55&0.63&2.77&13&483\\
1.3$\times$10$^{6}$ V/m (up)  &7.61&0.54&0.61&2.52&13&439\\
0.8$\times$10$^{6}$ V/m (up)  &7.28&0.55&0.60&2.38&15&405\\
0                  V/m        &7.15&0.55&0.60&2.35&16&431\\
0.8$\times$10$^{6}$ V/m (down)    &7.10&0.55&0.59&2.33&16&376\\
1.3$\times$10$^{6}$ V/m (down)    &6.78&0.55&0.61&2.30&14&451\\
1.7$\times$10$^{6}$ V/m (down)    &6.60&0.55&0.58&2.15&19&407\\
2.1$\times$10$^{6}$ V/m (down)    &6.50&0.55&0.59&2.10&18&428\\
\noalign{\smallskip}\hline
\hline\noalign{\smallskip}
\end{tabular}
\end{table*}

\begin{figure*}[htbp]
\centering
\includegraphics[width=1\textwidth]{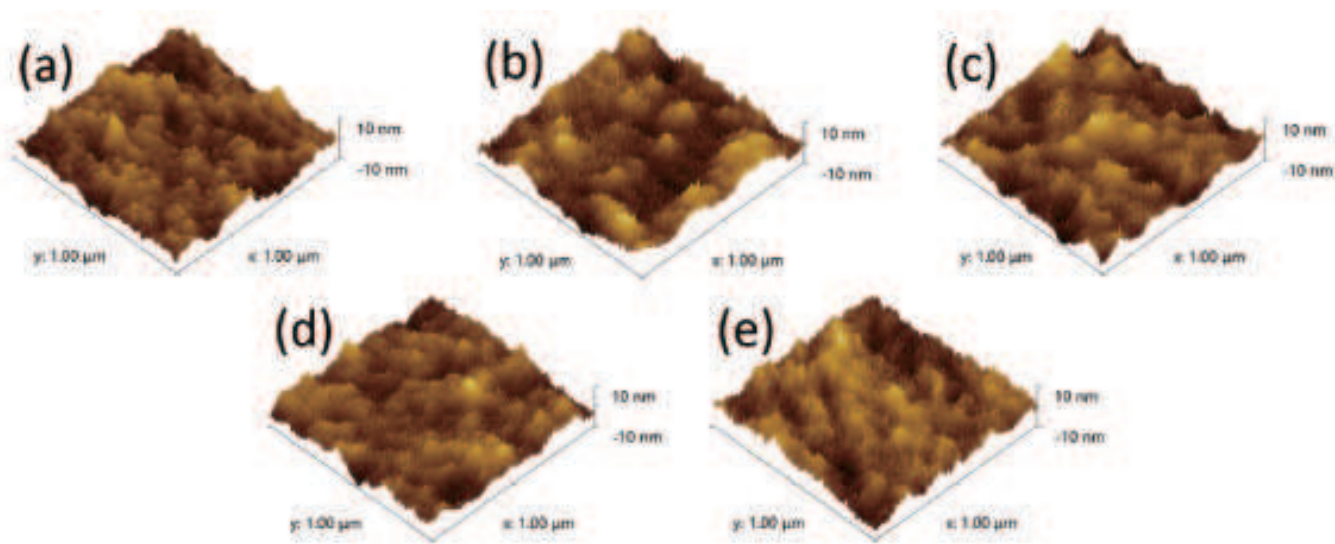}\hspace{0.05in}

\caption{AFM height images of the surface of P3HT:PCBM thin films (a) with no EEF treatment, and applied with the vertical EEF for (b) +1.7$\times$10$^{6}$ V/m, (c) +2.1$\times$10$^{6}$ V/m, (d) -1.7$\times$10$^{6}$ V/m, (e) -2.1$\times$10$^{6}$ V/m. }

\label{F3}       
\end{figure*}

\end{document}